\documentclass[aps,prb, twocolumn, superscriptaddress, floatfix,10pt]{revtex4}

\usepackage{graphicx}
\usepackage{dcolumn}
\usepackage{bm}

\begin{document}

\title{Single Shot Charge Detection Using A Radio-Frequency Quantum Point Contact}
\author{M. C. Cassidy\footnotemark[2]}
\author{A. S. Dzurak}
\author{R. G. Clark}
\affiliation{Centre for Quantum Computer Technology, Schools of Electrical Engineering and Physics, University of New South Wales, Sydney 2052, Australia}
\author{K. D. Petersson\footnote[2]{These authors contributed equally to this work.}}
\author{I. Farrer}
\author{D. A. Ritchie}
\author{C. G. Smith}
 \affiliation{Cavendish Laboratory, JJ Thomson Road, Cambridge CB3 0HE, United Kingdom}

\begin{abstract}
We report on charge sensing measurements of a GaAs semiconductor quantum dot device using a radio frequency quantum point contact (rf-QPC). The rf-QPC is fully characterized at 4~K and milli-Kelvin temperatures and found to have a bandwidth exceeding 20 MHz. For single-shot charge sensing we achieve a charge sensitivity of $\sim2 \times10^{-4} e/\sqrt{Hz}$ referred to the neighboring dot's charge. The rf-QPC compares favorably with rf-SET electrometers and promises to be an extremely useful tool for characterizing and measuring semiconductor quantum systems on fast timescales.
\end{abstract}

\maketitle

Recently there has been considerable interest in using non-invasive charge detectors to probe semiconductor quantum dots, due to their potential applications in quantum information processing \cite{Petta2005, Elzerman2004, Gustavsson2006}. These charge detection schemes have primarily been based around the use of a quantum point contact (QPC) capacitively coupled to a neighboring quantum dot device \cite{Field1993}. When biased at around $G_0 = e^2/h \approx 38.7 \mu S$, midway between the first conductance plateau and pinch-off, the QPC is extremely sensitive to variations in the local electrostatic environment. However, when using DC measurement techniques the best reported bandwidths for real-time charge sensing have been $\leq 100~kHz$ which have been limited by the line capacitance and $1/f$ noise \cite{Meunier2007, Vandersypen2004}.

The application of radio frequency resonant circuit techniques to single electron devices has led to the development of an extremely fast and sensitive electrometer known as a radio-frequency single electron transistor (rf-SET) \cite{Schoelkopf1998}. rf-SETs have been coupled to a variety of structures, including semiconductor quantum dots with impressive results \cite{Lu2003}, however they introduce an extra level of complexity in the fabrication of a device. Like the SET, QPCs have a very low barrier capacitance, $\sim$ 1 fF \cite{Regul2004}, giving them intrinsically high bandwidths of several gigahertz and should therefore be amenable to similar high frequency techniques. In this letter, we demonstrate how the performance and bandwidth of an rf-SET can be combined with the simplicity of fabrication of a QPC to form an rf-QPC with which we carry out fast charge sensing.  This builds on previous work which identified the feasibility of an rf-QPC as a fast and sensitive electrometer \cite{Qin2006}. Two rf-QPCs were measured at milli-Kelvin temperatures, both showing similar results. Here we present the results for one of the measured rf-QPCs, which was coupled to a double quantum dot device.
The device was fabricated using a GaAs/AlGaAs heterostructure with a high mobility two dimensional electron gas (2DEG) 97nm below the surface. The ungated 2DEG had a mobility at 1.3K of $1.36\times10^6 cm^2/Vs$ and a carrier concentration in the dark of $1.36\times10^{11} cm^{-2}$. The quantum dots and point contact were formed in the 2DEG using reverse-biased Ti/Au Schottky gates. A SEM micrograph of the device layout is shown in Figure 1(a).  All measurements, unless otherwise stated, were carried out in a dilution refrigerator with a base temperature of $\sim60~mK$.

\begin{figure}
		\includegraphics[scale=1]{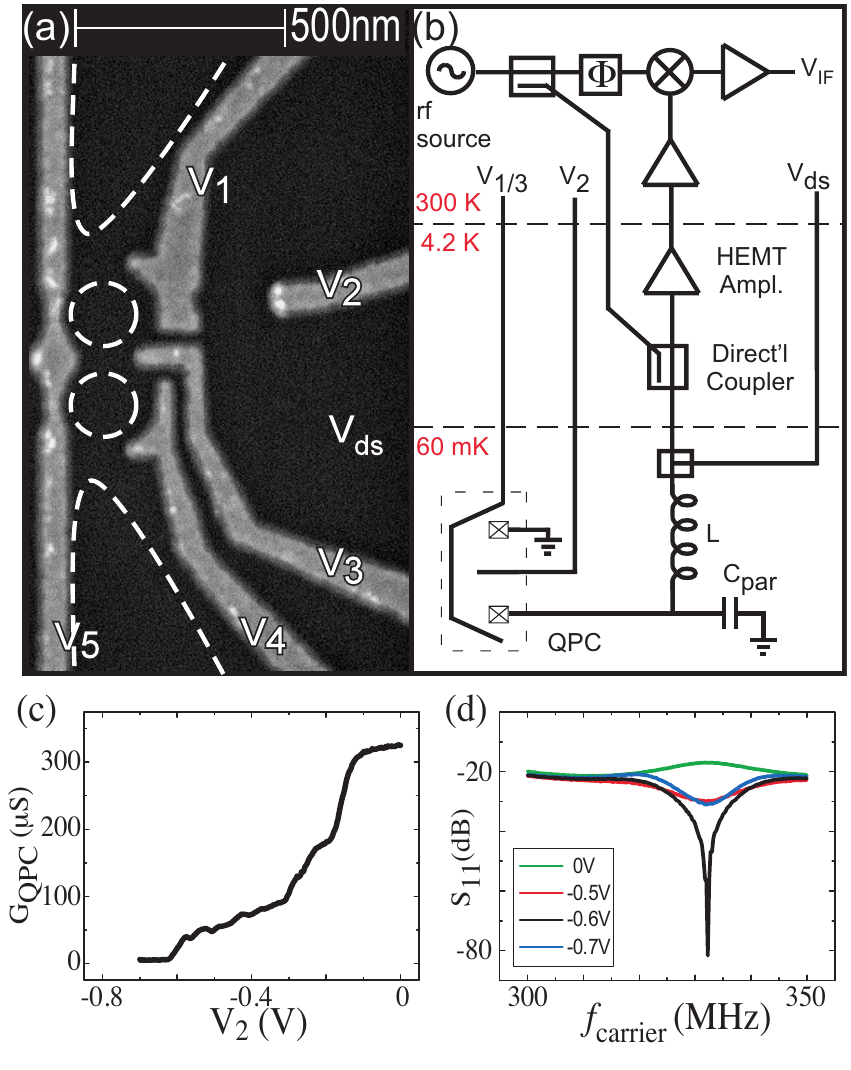}
	\caption{\label{fig:epsart} (a) Scanning electron micrograph of a device similar to the one measured. (b) A simplified schematic of the measurement circuit for the rf-QPC. (c) DC conductance vs QPC gate voltage. (d) RF reflected power vs carrier frequency and QPC gate voltage. } 
\end{figure}

The quantum point contact was embedded in a resonant tank circuit as shown in Figure 1(b). The carrier signal was coupled to the sample using a directional coupler with the reflected signal then fed into cryogenic and room-temperature amplifiers, both with gains of $40~dB$. For time-domain measurements the signal was demodulated using a homodyne detection technique \cite{Buehler2004} and then further amplified and filtered for sampling. For impedance matching, the inductor value ($L=490~nH$) was chosen so that it formed a resonant circuit with the stray pad capacitance ($C_{par}=0.47~pF$) at $f_0 \approx 332~MHz$ which was within the $300-400~MHz$ noise-temperature bandwidth of our cold amplifier. We were able to simultaneously monitor the QPC conductance to ground using a custom-built bias-tee. A low-frequency excitation (113~Hz) of $100~\mu V$ ($10\mu V$ for time-domain measurements) was applied to the source and the voltage drop across a $2~k\Omega$ series resistance was then measured using lock-in detection. dc conductance and rf reflected power versus QPC gate voltage are shown in Figure 1(c)-(d) . The tank circuit achieved matching around a conductance of $G_{QPC} = 30~\mu~S$  which is close to the optimal operating point of $G_0$ for charge sensing.

We characterized the rf-QPC using gate voltage values equivalent to a charge movement in the  quantum dot \cite{Perez-Martinez2007}. A small voltage oscillation of 0.35~mV or 1.4~mV was applied to gates $V_{2}$ and $V_{4}$ respectively and the signal to noise ratio (SNR) of the reflected signal measured. The sensitivity of the rf-QPC is then measured in terms of the noise voltage $ = \Delta V / \sqrt{B}10^{\frac{SNR}{20}}$. Figure 2(a) shows the sensitivity of the rf-QPC as a function of to the local gate oscillation $V_{2}$. The sensitivity of the rf-QPC was greatest between the first plateau and pinch-off. The sensitivity was also found to be slightly better for a carrier signal just off the resonance frequency at $326~MHz$. Figure 2(b) shows the frequency response of our rf-QPC which has a 3~dB bandwidth of $\sim21~MHz$ giving a quality factor of $Q \approx 8$. $1/f$ noise is also significant up to $\sim100~kHz$ which is similar to reported values for Al rf-SETs \cite{Buehler2005}. Figure 2(c) shows that maximum sensitivity is for a carrier power (at the sample) of around $-60dBm$.  At $4~K$, for sufficiently high carrier powers, the sensitivity is only a factor of two worse, demonstrating that the rf-QPC can serve as an effective change detector at 4~K. 

\begin{figure}
\includegraphics[scale=1]{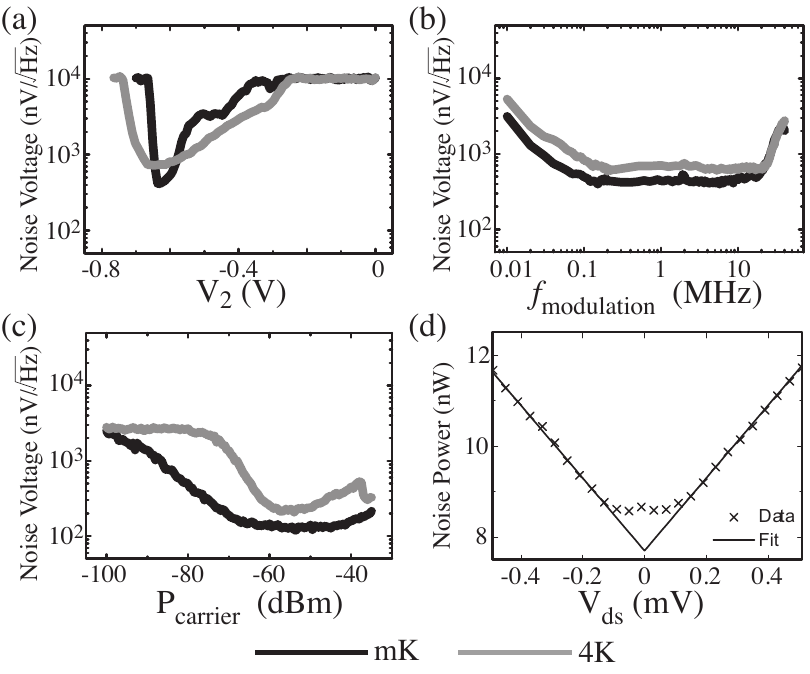}
\caption{\label{fig:epsart} Sensitivity measurements for the rf-QPC at 4K and mK as a function of different parameters: (a) the rf-QPC gate voltage; (b) the frequency of the gate modulation; and (c) the power of the carrier source at the sample. A lower noise voltage corresponds to a higher sensitivity. In (a) and (b) $V_4$ was modulated while for (c) $V_2$ was modulated. (d) Amplifier noise power measurements at mK temperatures with a fit according to Eq.(1)}  
\end{figure}

To analyze the noise contribution of the amplifier and extract the system noise temperature, $T_n$, we plot the noise power (at $326~MHz$) as a function of source-drain bias, $v_{ds}$, in Figure 2(d). Assuming matching, for $|v_{ds}| > k_B T/e$ the noise power will tend to \cite{Aassime2001}:
\begin{eqnarray} 
P = AB(k_B T_n + \frac{1}{2} e \eta |v_{ds}|) 
\end{eqnarray}
With $B = 1~MHz$ and a Fano factor of $\eta < 1$\cite{Kumar1996} we extract an upper bound for our noise temperature of $T_N < 5.8~K$ (with a gain of $A = 80~dB$). We note that, as the 1D subband spacing of a QPC is larger than the typical charging energy for a SET, it is possible to operate the rf-QPC at higher source-drain biases making the system noise contribution to the total noise less significant.

As the capacitance between the gate and the rf-QPC cannot be measured we are unable to extract a charge sensitivity figure that is equivalent to those stated for a rf-SET. We therefore turn our attention to actual charge sensing experiments using a double quantum dot device coupled closely to our rf-QPC. In this case, the charge sensitivity is dependent on the rf-QPC/device coupling. 

For the following charge sensing measurements we have operated the device as effectively one large dot by setting the gate voltage on the central barrier at a small negative voltage. Figures 3(a) show a map of a section of the device charge stability diagram when the device is operated as a single dot, showing the derivative of the demodulated rf-QPC signal (with respect to $V_{4}$) as a function of both $V_{1}$ and $V_{4}$. The charge stability diagram was built up by applying a $100 mV$, $2 kHz$ triangular wave to gate $V_{4}$ and averaging over 64 sweeps.  $V_{1}$ was then stepped and $V_{2}$ compensated to ensure that $G_{QPC}$ was close to its optimal value for charge sensing. 

\begin{figure}
\includegraphics[scale=0.25]{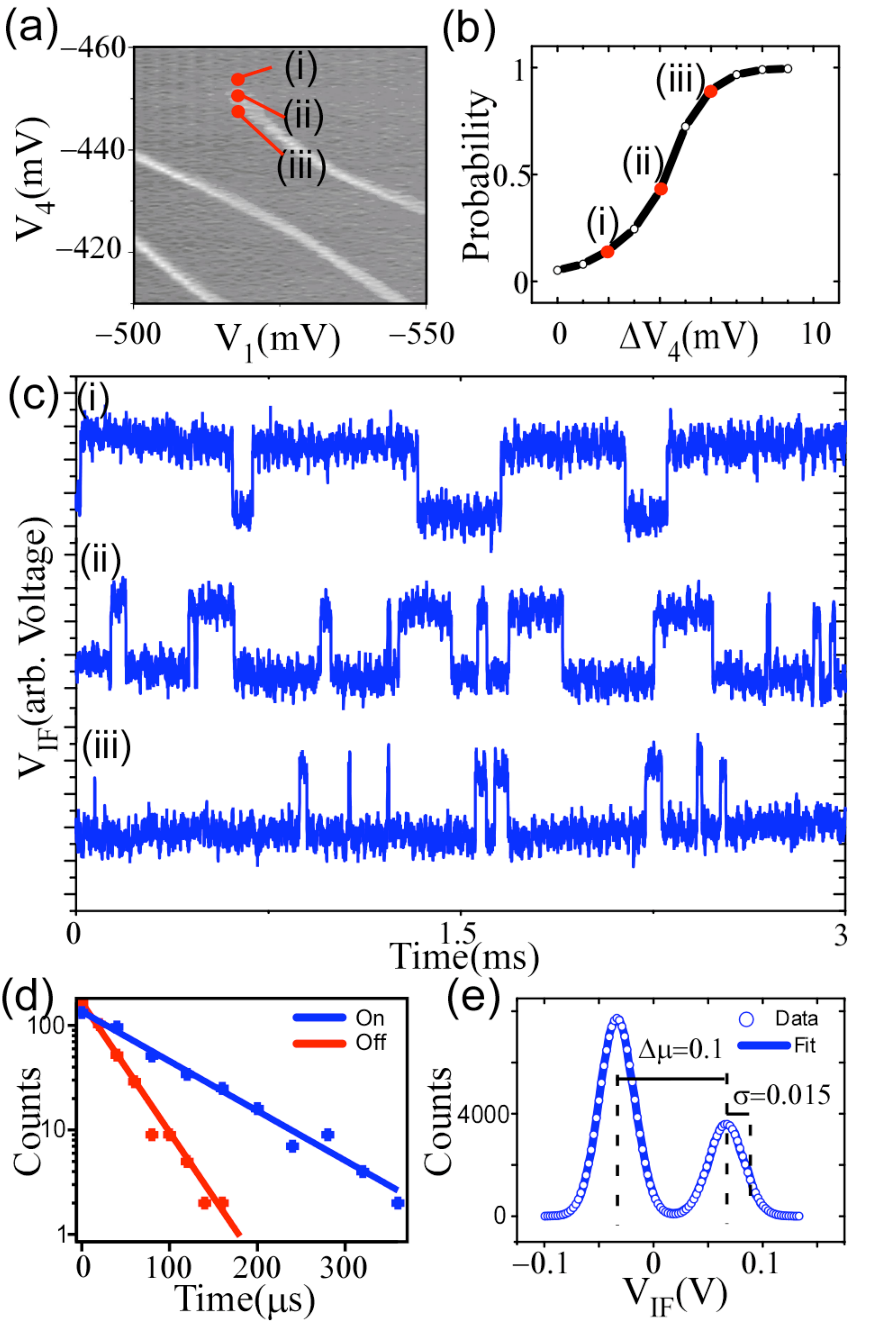}
\caption{\label{fig:epsart} (a) Charge stability diagram for the device when operated as a single dot measured using the rf-QPC. Each vertical sweep is the derivative of the rf-QPC signal averaged over 64 cycles. Points (i)-(iii) indicates the operating points for measurements in (c). (b) The electron occupation probability of the quantum dot as $V_{4}$ is swept across the charge degeneracy point from (i)-(iii).(c) Single shot traces for each of the points (i)-(iii) in (a). (d) Histograms for the tunneling times off and on the dot at a particular point close the charge degeneracy.
(e) The histograms of a long time-dependent trace fitted to a double Gaussian distribution.}
\end{figure}

These charge sensing measurements were carried out using an rf-power of $-73~dBm$ at the sample which, based on Figure 2(e), is less than required for optimal charge sensitivity. Assuming the tank circuit is impedance matched to $Z_0 = 50~\Omega$, the source-drain voltage across the QPC is given by $ v_{ds} \approx Q_{e} v_{rf}$ where $v_{rf}$ is the open circuit voltage of the rf source. For a $-73~dBm$ carrier signal this corresponds to an excitation of $v_{ds} \approx 1~mV$.  Going to higher powers, for example $-63~dBm$ ($v_{ds} \approx 3~mV$), the differentiated charge detection signal split into double peaks indicating that the oscillating rf-QPC drain was modulating the dot energy via the capacitive coupling. We note that this problem does not occur for conventional charge sensing as the drain is fixed at a constant bias, however, it has been observed that even for a constant bias $v_{ds} > 1~mV$ the tunneling statistics of the dot are significantly altered \cite{Vandersypen2004}. 

At more negative voltages of the two gates, $V_{1}$ and $V_{4}$, the charge transition lines can be seen to disappear, indicating that the tunneling rate is becoming comparable to the sweep rate. To observe real-time tunneling events on and off the dot we biased the dot in this region around point (ii) in Figure 3(a). Figure 3(b) shows the occupation probability as a function of gate voltage, sweeping across the charge degeneracy point. Figure 3(c)(i-iii) shows a typical detector trace taken primarily in the `N' electron state (i), near equilibrium (ii) and in the `N+1' electron state (iii) with a detection bandwidth of $B \approx 500~kHz$  \cite{SRS560}.  Figure 3(d) gives histograms of tunneling times on and off the dot at a particular operating point which show the exponential decay expected for a Poissonian process \cite{Gustavsson2006}.

For a long time-dependent trace at a fixed operating point, we can form a histogram of the rf-QPC signal. Fitting this data to a double Gaussian, the RMS noise amplitude is then given by the standard deviation, $\sigma$, and the difference between the two means, $\Delta \mu$ gives the signal amplitude for a single-electron charging event. Our charge sensitivity is then given by:
\begin{equation} \delta q =  \frac{ 1 }{ \sqrt{B} \frac{ \Delta \mu}{ \sigma} } (e/ \sqrt{Hz})
\end{equation}
Based on Figure 3(e) we determine our charge sensitivity to be $\delta q \approx 2 \times 10^{-4} e/\sqrt{Hz}$ referred to the quantum dot charge. Comparing with similar real-time data for an rf-SET coupled to a GaAs quantum dot \cite{Lu2003} we find that the sensitivities are similar (we estimate their charge sensitivity to be $\delta q \approx 3 \times10^{-4} e/\sqrt{Hz}$ referred to the quantum dot charge). We now consider how much better the charge sensitivity could be. Assuming that the sensitivity is limited by shot noise ($S_I = 2e \eta G v_{ds}$) the charge sensitivity is given by \cite{Grabert1992}:
\begin{eqnarray} 
\delta q =& \frac{1}{|dI/dq|}\sqrt{S_I}    \\
=& |\frac{\Delta q}{\Delta G}|\sqrt{\frac{2e \eta G}{v_{ds}}}
\end{eqnarray}
For a similar device, the dot-to-QPC coupling was measured to be $ |\frac{\Delta q}{\Delta G}| \approx e / 0.05 G_0 $ \cite{Perez-Martinez2007}. Taking then $ G=G_0 $, $ \eta = 0.5$ and $v_{ds} = 1~mV$ we get a lower bound of $\delta q  \approx 4 \times 10^{-5} e/\sqrt{Hz}$. Qualitatively, we attribute this discrepancy to system noise, additional demodulation circuit noise and to a possibly weaker dot-to-QPC coupling constant than the value assumed.

In summary, we have demonstrated charge sensing of a quantum dot device using an rf-QPC. We estimate a time-domain charge sensitivity of $\delta q \approx2\times10^{-4} e/\sqrt{Hz}$ referred to the dot charge which is comparable to sensitivities obtained using rf-SETs. Thus the rf-QPC combines simplicity of fabrication with excellent sensitivities even at 4K. Using our rf-QPC we have been able to rapidly characterize our quantum dot device and observe real-time tunneling on the time scale of less than 5 microseconds. 

The authors wish to thank A.J. Ferguson and T. Duty for helpful discussions and D. Anderson, R.P. Starrett and D. Barber for technical support. This work is supported by the EPSRC, the Australian Research Council under Linkage Grant LX0669069, the Australian Government and by the US National Security Agency (NSA) and US Army Research Office(ARO) under Contract No. DAAD19-01-1-0653. K.D.P. acknowledges the support of Trinity College, Cambridge.

\newpage


\begin{thebibliography}{99}
\bibitem{SRS560} This measured 3dB bandwidth of $500 kHz$ was set by our Stanford Research SR560 pre-amplifier with a 1MHz filter cut-off and 12 dB/octave roll-off.
\end{thebibliography}
\end{document}